\documentclass[a4paper,11pt]{article}
\pdfoutput=1 

\usepackage{jheppub} 

\usepackage[T1]{fontenc} 

\title{\boldmath 
Fate of the first-order chiral phase transition in QCD:
Implications for dark QCD studied via a Nambu–Jona-Lasinio model 
}

\author[a]{Yuanyuan Wang,}
\author[b]{Mamiya Kawaguchi,}
\author[a]{Shinya Matsuzaki,}
\author[c]{and Akio Tomiya}

\affiliation[a]{Center for Theoretical Physics and College of Physics, Jilin University, Changchun, 130012,
China}
\affiliation[b]{School of Nuclear Science and Technology, University of Chinese Academy of Sciences, Beijing
100049, China}

\affiliation[c]{Department of Information Technology, International Professional University of Technology in Osaka, 3-3-1 Umeda, Kita-Ku, Osaka, 530-0001, Japan}

\emailAdd{yuanyuanw20@mails.jlu.edu.cn}
\emailAdd{mamiya@ucas.ac.cn}
\emailAdd{synya@jlu.edu.cn}
\emailAdd{akio@yukawa.kyoto-u.ac.jp}

\abstract{
The first-order nature of the chiral phase transition in QCD-like theories 
can play crucial roles to address a dark side of the Universe, 
where the created out-of equilibrium is essential to serve as 
cosmological and astrophysical probes such as gravitational wave productions, 
which have extensively been explored. 
This interdisciplinary physics is built based on 
a widely-accepted conjecture that 
the thermal chiral phase transition in QCD-like theories with massless (light) three flavors is of first order. 
We find that such a first order feature may not hold, 
when ordinary or dark quarks are externally coupled to a weak enough background field of photon or dark photon (which we collectively call a ``magnetic" field). 
We assume that a weak ``magnetic" background field could be 
originated from some ``magnetogenesis" in the early Universe. 
We work on a \textcolor{black}{Nambu-Jona-Lasinio model} which can describe the chiral phase transition in a wide class of QCD-like theories.  
We show that in the case with massless (light) three flavors, 
the first-order feature goes away when $2 f_\pi^2 \lesssim eB ( \ll (4 \pi f_\pi)^2)$, where $eB$ is the ``magnetic" field strength and  
$f_\pi$ the pion decay constant at the vacuum. 
This disappearance is the generic consequence of the presence 
of the ``magnetically" induced scale anomaly and the ``magnetic" catalysis for the chiral symmetry 
breaking, and would impact or constrain modeling dark QCD coupled to an external ``magnetic" field. 
}

\begin{document} 
\maketitle
\flushbottom

\section{Introduction}
\label{sec:intro}

Pursuing the nature of the chiral phase transition in thermal QCD 
is crucial to deeply comprehend  
the origin of matter mass in light of the thermal history of the Universe. 
This research direction has extensively been explored since 
the pioneer work by Pisarski and Wilczek~\cite{Pisarski-Wilczek}, where 
the analysis was done in the 
chiral limit for the lightest three or two flavors based on 
the renormalization group analysis by employing a chiral effective model for 
low-energy QCD. 
The remarkable extension was made by the Columbia university group  
utilizing lattice QCD calculations~\cite{Columbia}, which has been 
refined along with development in modern computational technologies. 
In lattice QCD, 
the QCD phase transition including the chiral 
and confinement-deconfinement phase transitions 
have been investigated 
taking into the quark mass and the number of flavor dependence.  
It is summarized as the so-called Columbia plot, 
which is drawn on the quark mass plane with 
the classification by the universality classes  
in terms of the renormalization group.

Currently, the interest has been extended from the original motivation of deeply understanding the QCD phase transition to applications to Beyond the Standard Models, which has opened a vaster market involving astrophysical and cosmological fields.  
Of late fashion, 
a dark side of the Universe is discussed by assuming what is called dark or hidden QCD which can leave various cosmological and astrophysical footprints in the present-day Universe, such as gravitational wave spectra, 
dark MACHOs (Massive Astrophysical Compact Halo Objects) including 
dark compact stars, and dark matter abundances. 
Then, QCD with light or massless three flavors 
is often referred to as a typical and promising 
candidate which can realize the first-order chiral phase transition to supply 
such footprints~\cite{Holthausen:2013ota,Ametani:2015jla,Aoki:2017aws,Bai:2018dxf,Helmboldt:2019pan,Reichert:2021cvs,Archer-Smith:2019gzq,Aoki:2019mlt,Dvali:2019ewm,Easa:2022vcw,Tsumura:2017knk}.

This trend has been inspired by the pioneer work~\cite{Pisarski-Wilczek} 
showing the possibility of the first-order chiral phase transition for 
the number of flavors greater than two, though it is based on a perturbative 
analysis within the framework of a chiral effective model (linear sigma model). 
Actually, no clear lattice results on the full QCD simulations for the first-order phase-transition has been obtained, and instead, only lower limits on the quark mass in the crossover regime has been placed~\cite{3F-HISQ, 3F-Wilson}.  
Thus the first-order nature has not been conclusive yet.


Besides, the early hot Universe could have been more intricate: 
it might have been covered with space-homogeneous or constant  
background fields of photon or others such as those associated with a dark photon.  
Those background fields, represented as a primordial ``magnetic" field, could be produced via cosmological first-order phase transitions~\cite{Vachaspati:1991nm,Enqvist:1993np,Grasso:1997nx,Grasso:2000wj,Ellis:2019tjf,Zhang:2019vsb,Di:2020ivg,Yang:2021uid}, 
or inflationary scenarios~\cite{Turner:1987vd,Garretson:1992vt,Anber:2006xt,Domcke:2019mnd,Domcke:2019qmm,Patel:2019isj,Domcke:2020zez,Shtanov:2020gjp,Okano:2020uyr,Cado:2021bia,Kushwaha:2021csq,Gorbar:2021rlt,Gorbar:2021zlr,Gorbar:2021ajq,Fujita:2022fwc}, 
to be redshifted to the later epoch in which the Universe undergoes 
the ordinary QCD or dark QCD phase transition.  
Thereby, such hot QCD mediums might be weakly ``magnetized" by redshifting from the primordial 
``magnetogenesis", so that the chiral phase transition might take place in the 
thermal bath with a weak background ``magnetic" field to which quarks or dark quarks couple.

In the scenario of dark QCD, allowing 
dark quarks to couple with a dark photon would make it possible to 
address the dark pion as 
a dark matter candidate and produce its cosmological abundance due to    
introduction of the kinetic term mixing between the photon and dark photon, as discussed in the context of SIMPs (Strongly Interacting Massive Particles)~\cite{Hochberg:2014dra,Hochberg:2014kqa,Hochberg:2015vrg}. Therefore, the background ``magnetic" field would be favored and naturally present in dark QCD models in a sense of the phenomenological and cosmological interests,  
as well as ordinary QCD.

The possible presence of a weak background ``magnetic" field has naively been disregarded in addressing the order of the chiral 
phase transition. However, it turns out to be too naive.

In this paper, we find that  
when ordinary or dark quarks are externally coupled to a weak enough background field of photon or dark photon (which we hereafter call a ``magnetic" field collectively), thermal QCD with light three flavors may not undergo the first-order chiral phase transition.

To make an explicit demonstration, we work on an Nambu-Jona-Lasinio (NJL) model 
which can describe the chiral phase transition in 
a wide class of QCD-like theories including dark QCD. 
We include the contribution from 
the scale anomaly induced by quark couplings to a ``magnetic" field, as discussed 
in the literature~\cite{Kawaguchi:2021nsa}, 
which is present only in a weak field regime. 
\textcolor{black}{
It then turns out that the disappearance of the first-order nature 
in light three-flavor QCD 
arises as 
the generic consequence of the crucial presence 
of the ``magnetically" -induced scale anomaly and the ``magnetic"  catalysis~\cite{Klevansky:1989vi,Klimenko:1990rh,Shovkovy:2012zn} for the chiral symmetry breaking at around the criticality, 
which is irrespective to details of the parameter setting in the NJL 
model description monitoring ordinary QCD. 
}

The size of the critical ``magnetic" field strength $(eB_{\rm cr})$ normalized to $f_\pi$ is estimated as $eB_{\rm cr}/f_\pi^2 \sim 2$, above  
which (but still at weak $eB$ $\ll (4 \pi f_\pi)^2$) the first-order feature goes away, 
where $f_\pi$ is the pion decay constant at the vacuum with $T=eB=0$.  
\textcolor{black}{
Further allowing quarks to have finite masses, we show 
that the first-order feature in the light quark mass regime 
is completely gone at the critical field strength, 
which is captured in a couple of extended Columbia plots including 
the $eB$-axis. 
} 

Our finding not merely helps deeply understanding the chiral phase structure in QCD, 
\textcolor{black}{but would thus impact modeling dark QCD in the NJL description and testing the model via the astrophysical probes and constrain the ``magnetogenesis"}.

\section{NJL model with electromagnetic scale anomaly}

We begin by briefly introducing an NJL model with the light three flavors as a low-energy effective model of QCD, and present the Lagrangian and the thermodynamic potential with a constant magnetic field (which we shall hereafter call the thermomagnetic potential) containing the contribution of the electromagnetic scale anomaly.

Throughout the present paper, 
we refer to the ordinary up, down, and strange quarks with 
the electromagnetic charges $2/3, -1/3$, and $-1/3$ in unit of the electric charge 
$e$ and adopt ordinary QCD 
as the monitor for a wide class of QCD-like theories. 
However, as will be clarified later, the qualitative features associated with the chiral 
phase transition are substantially intact even if 
the charge assignments and 
model-parameter sets are taken to be adjusted to some dark QCD setup 
with photon or dark photon, 
differently from the ordinary QCD case.

\subsection{Model description and parameter setting monitoring ordinary QCD}

The three-flavor NJL model externally coupled to a background photon field 
is described by the following Lagrangian~\footnote{For reviews on the three-flavor NJL model, see, e.g., Refs.~\cite{Hatsuda:1994pi,Rehberg:1995kh}.}: 
\begin{equation}
		\begin{aligned}
			\mathcal{L}=\bar{\psi} \left(i \gamma^{\mu} D_{\mu}- {\bf m}\right) \psi &+ \sum_{a=0}^8 G\left\{\left(\bar{\psi} \lambda^{a} \psi\right)^{2}+\left(\bar{\psi} i \gamma^{5} \lambda^{a} \psi\right)^{2}\right\} \\
			&-K\left[\operatorname{det} \bar{\psi}\left(1+\gamma_{5}\right) \psi+\operatorname{det} \bar{\psi}\left(1-\gamma_{5}\right) \psi\right]\,, 
		\end{aligned}
		\label{Lag:NJL}
	\end{equation}
where $\psi=(u, d, s)^{T}$ represents the Dirac fermion field forming the flavor SU(3) triplet of up, down, and strange quarks (with QCD color and Dirac spinor indices suppressed); ${\bf m}$ denotes the quark mass matrix 
${\bf m} = 
	{\rm diag} \{ m_0, m_0, m_s\}$;  
$\lambda^{a}(a=0, \cdots, 8)$ are the Gell-Mann matrices in the flavor space, and $\lambda^{0}=\sqrt{2 / 3} \cdot {\bf 1}_{3 \times 3}$. 
In Eq.(\ref{Lag:NJL}) 
the covariant derivative $D_{\mu}=\partial_{\mu}\cdot 1_{3\times 3} -i e \, Q_{\rm em} A_{\mu}$ contains the coupling term between the quark field and the external photon field $A_\mu$ along with the electromagnetic charges in $Q_{\rm em} = {\rm diag} \{q_u, q_d, q_s \}$, where $q_{u}=2/3$, $q_d= -1/3$, and $q_s = -1 /3$. 
For simplicity, we introduce a constant magnetic field along the $z$-direction in space to be embedded in the external photon field as $A_{\mu}=(0,0, B x, 0)$.

The four-fermion interaction with the coupling $G$ is invariant 
under the full chiral $U(3)_L\times U(3)_R$ symmetry with the chiral transformation: $\psi \to U \cdot \psi$ with $U= e^{- i \gamma_5 \sum_{a=0}^8  (\lambda^a/2) \theta^a }$ and the chiral phases $\theta^a$. 
This four-fermion operator is supposed to be induced from 
the one-gluon exchange process at the confinement scale.

	The Kobayashi-Maskawa-‘t Hooft (KMT) determinant term~\cite{Kobayashi:1970ji,Kobayashi:1971qz,tHooft:1976rip,tHooft:1976snw}
	with the coupling $K$ is a six-point interaction induced from the QCD instanton configuration coupled to quarks. 
	The KMT term preserves $SU(3)_L\times SU(3)_R$ invariance (associated with the chiral phases labeled as $a=1, \cdots, 8$), but breaks the $U(1)_A$ (corresponding to $a=0$) symmetry. 
	This interaction gives rise to the mixing between different flavors and also uplifts the $\eta^{\prime}$ mass to make $ \eta'$  no longer a Nambu-Goldstone boson.

	The quark mass terms with $m_f$ ($f=u,d,s$) and the couplings to the background photon field with the charges $e q_f$ are flavor dependent. 
	Therefore, those explicitly, but weakly enough, break the 
	chiral $U(3)_L\times U(3)_R$ symmetry as long as the strength of $eB$ is 
	perturbatively small, say, $eB \lesssim {\cal O}(f_\pi^2)$, which is indeed the case in the present setup. Since the $U(1)_A$ symmetry is to be badly broken by the KMT term so as to reproduce the observed $\eta'$ mass, 
	the model keeps the approximate chiral $SU(3)_L \times SU(3)_R$ symmetry only.

When the couplings $G$ and/or $K$ get strong enough, the approximate chiral $SU(3)_L \times SU(3)_R$ symmetry is spontaneously broken 
by nonperturbatively developed  
nonzero quark condensates $\langle \bar{\psi}_f \psi_f \rangle \neq 0$,
down 
to the vectorial symmetry $SU(3)_V$ (corresponding to invairance associated with 
the vectorial phase rotation $\psi \to e^{- i \sum_{a=0}^8  (\lambda^a/2) \theta^a } \cdot \psi$), to be consistent with underlying QCD. 
The present NJL model monitors the spontaneous breakdown by the large $N_c$ expansion or equivalently in the mean field approximation, 
where $N_c$ stands for the number of QCD colors, to be set to 3.  

\textcolor{black}{
    The NJL model itself is nonrenormalizable because the $G$ and $K$ coupling terms are higher dimensional operators greater than four in mass dimension.  
    Therefore, a momentum cutoff $\Lambda$ needs to be introduced to make the NJL model regularized.} 

    Relevant NJL formulas to compute QCD hadronic observables at the vacuum are presented in Appendix~\ref{App:formulas}. 
	We quote the literature~\cite{Rehberg:1995kh} \textcolor{black}{where the two coupling parameters $G$ and $K$ normalized to a three-dimensional momentum cutoff $\Lambda$ are fixed as}  
	\begin{align} 
	G\Lambda^2 \simeq 1.835, \qquad K\Lambda^5 \simeq 12.36
\,, \label{parameters1}
	\end{align}
	with $\Lambda$ to the pion decay constant $f_\pi$:  
	\begin{align} 
	\Lambda/f_\pi \simeq  6.518 
	\,. \label{parameters2}
	\end{align}
	This parameter set has been determined by 
	fixing four representative hadronic observables in the isospin-symmetric limit at $T=eB=0$, such as 
	the pion decay constant $f_\pi = 92.4$ MeV, 
	the pion mass $m_\pi \simeq  135.0$ MeV, the kaon mass $m_K \simeq 497.7$ MeV, 
	and the $\eta'$ mass $m_{\eta'} \simeq 957.8$ MeV, with $m_0=5.5$ MeV input, which also yields $m_s = 140.7$ MeV,   $\Lambda=602.3$ MeV, and the $\eta$ mass $m_\eta \simeq 514.8$ MeV. 
	Those isospin-symmetric hadronic quantities are in good agreement with the recent lattice results at 
	the physical point of 2 + 1 flavor QCD with 
	$m_\pi = 138$ MeV~\cite{Aoki:2013ldr}:  
	$m_K =(494.2 \pm 0.4)$ MeV~\cite{Aoki:2013ldr}, 
	$m_\eta = (554.7 \pm 9.2)$ MeV~\cite{Bali:2021qem}, 
	and $m_{\eta'} = (930 \pm 21)$ MeV~\cite{Bali:2021qem}.

	When working at finite temperature, we will not consider intrinsic-temperature dependent couplings, 
instead, all the $T$ dependence should be induced only 
from the thermal quark loop corrections to the couplings 
defined and introduced at the vacuum with $T=eB=0$. 
Actually, the present NJL at $eB=0$ qualitatively fits well with lattice QCD results at the physical point on the temperature dependence 
for the chiral, axial, and topological susceptibilities, 
as shown in Ref.~\cite{Cui:2021bqf}~\footnote{
The values of the fixed model parameters in the literature~\cite{Cui:2021bqf} are slightly different from those in the present analysis, which, however, only gives tiny discrepancy by  about a few percent level in evaluating the $T$ dependence of the chiral order parameter. 
}. 
In this sense, we do not need to introduce such an additional $T$ dependence into  the model parameters.

The QCD-monitored parameter setup in Eqs.(\ref{parameters1}) and  (\ref{parameters2}) forms a wide class of QCD-like theories in the NJL description obtained by scaling up of ordinary QCD with respect to $\Lambda$ or $f_\pi$. 
\textcolor{black}{
A similar parameter setup derived based on a different regularization scheme
has been applied in dark QCD scenarios with use of the scaling up~\cite{Holthausen:2013ota,Ametani:2015jla,Aoki:2017aws,Reichert:2021cvs,Helmboldt:2019pan,Aoki:2019mlt}.}


\subsection{Thermomagnetic potential with electromagnetic scale anomaly}

We work in the mean field approximation, so that 
from Eq.(\ref{Lag:NJL}) 
the constituent quark masses (i.e. the full quark masses) are read off as~\cite{Rehberg:1995kh}
	\begin{align}
		M_{u} &=m_{u}-4 G \phi_{u}+2 K \phi_{d} \phi_{s} \equiv m_0 + \sigma_u, 
	 \notag\\ 
	 M_{d} &=m_{d}-4 G \phi_{d}+2 K \phi_{u} \phi_{s} \equiv m_0 + \sigma_d, 
	 \notag\\ 
	 M_{s} &=m_{s}-4 G \phi_{s}+2 K \phi_{u} \phi_{d} \equiv m_s + \sigma_s, 
	 \notag\\ 
{\rm with}	\qquad 
\phi_{f} &\equiv \langle \bar{q}_f q_f  \rangle =  - i N_c \operatorname{tr} \int\frac{d^4p}{(2\pi)^4}S^{f}(p)	
	\,,\label{Mfs}
	\end{align}
where 
$\sigma_f$s correspond to the dynamical quark masses, 
and $S^{f}(p)$ denotes the $f$-quark full propagator in the mean field approximation, which, in the Landau level representation, takes the form~\cite{Miransky:2015ava}
\begin{align} 
S^f(p) = i e^{-p_{\perp}^{2} l^{2}_{f}} \sum_{n=0}^{\infty} \frac{(-1)^{n} D_{n}^f(p)}{p_{0}^{2}-p_{3}^{2}-M^{2}_f-2 n\left|q_{f}  e B\right|} 
\,, \label{Sf}
\end{align}
with $p_\perp \equiv (p_1, p_2)$, $l_f \equiv \sqrt{1 /\left|q_{f} e B\right|}$,  
and 
\begin{equation}
\begin{aligned}
D_{n}^f(p)=& 2\left[p^{0} \gamma^{0}-p^{3} \gamma^{3}+M_f \right] \\
& \times\left[\mathcal{P}_{\perp}^f L_{n}\left(2 p_{\perp}^{2} l^{2}_f\right)-\mathcal{P}_{-}^f L_{n-1}\left(2 p_{\perp}^{2} l^{2}_f \right)\right] \\
&+4\left(p^{1} \gamma^{1}+p^{2} \gamma^{2}\right) L_{n-1}^{1}\left(2 p_{\perp}^{2} l^{2}_f \right) 
\,. 
\end{aligned}
\end{equation}
Here $L_{n}^{\alpha}(z)$ are the generalized Laguerre polynomials, $\mathcal{P}_{\pm}^f \equiv \frac{1}{2}\left(1 \pm i s^f_{\perp} \gamma^{1} \gamma^{2}\right)$ with $s_{\perp}^f=$ $\operatorname{sign}\left(q_{f} e B\right)$.

To evaluate thermal and magnetic contributions to the thermomagnetic potential 
in the mean field approximation, we apply  the imaginary time formalism 
and the Landau level decomposition. 
The following replacements can then be activated~\footnote{ 
\textcolor{black}{ 
In the present study, 
we simply apply a conventional soft cutoff scheme~\cite{Frasca:2011zn} to 
regularize the integration over $p_3$ in Eq.(\ref{replacement}). 
In this regularization scheme, the 
parameter setting in Eqs.(\ref{parameters1}) 
and (\ref{parameters2}) together with $m_0=5.5$ MeV, $m_s = 140.7$ MeV, and $f_\pi = 92.4$ MeV yields $m_\pi \simeq 138$ MeV at $T=eB=0$, which differs by about 2 \% from the one ($m_\pi \simeq 135$ MeV)
based on a three-momentum 
cutoff scheme with the same inputs used [See the discussion below  Eq.(\ref{parameters2})]. 
The same level of discrepancy can be observed in other meson masses. 
This small dependence on the regularization scheme does not make substantial difference on the qualitative trend for the chiral-phase transition nature which we will mainly discuss later on. 
} 
}: 
\begin{equation}
\begin{aligned}
p_{0} & \leftrightarrow i \omega_{\mathbf{k}}=i(2 \mathbf{k}+1) \pi T, \\
\int \frac{d^{4} p}{(2 \pi)^{4}} & \leftrightarrow i T \sum_{\mathbf{k}=-\infty}^{\infty} \int \frac{d^{3} p}{(2 \pi)^{3}} \leftrightarrow i T \sum_{\mathbf{k}=-\infty}^{\infty} \sum_{f=u, d, s} \sum_{n=0}^{\infty} \frac{\left|q_{f} e B\right|}{2 \pi} \int_{-\infty}^{\infty} \frac{d p_{3}}{2 \pi}\,.
\end{aligned}
\label{replacement}
\end{equation}
Here $\omega_{\bf k}$ denotes the Matsubara frequency with the level of ${\bf k}$ and 
$n$ represents Landau levels.

\subsubsection{Magnetically induced scale anomaly terms}

As discussed in Ref.~\cite{Kawaguchi:2021nsa}, 
in the thermomagnetic medium with a weak enough magnetic 
field strength ($eB \ll (4 \pi f_\pi)^2$), 
the electromagnetic scale anomaly induces 
tadpole potential terms for the chiral singlet component of 
the scalar mesons. 
At the quark one-loop level, the induced potential takes the form
\begin{eqnarray}
{V}_{\rm eff}^{\rm (Tad)}(M_f, T, eB)
=
-
\frac{\varphi}{f_\varphi}
T^\mu_\mu(M_f, T, eB)
\,. \label{EM-scale-ano}
\end{eqnarray}
$\varphi$ denotes the chiral singlet part of the scalar mesons arising as the radial component of the nonet-quark bilinear $ \bar{q}_f q_{f'} $, so that 
$\varphi \approx \sqrt{\sigma_u^2 + \sigma_d^2 + \sigma_s^2}$, 
where $\sigma_f$s are diagonal elements of $ \bar{q}_f q_{f'} $. 
$f_\varphi$ is the vacuum expectation value of $\varphi$ at $T=eB=0$. 
Then the prefactor $\varphi/f_\varphi$ in Eq.(\ref{EM-scale-ano}) is given as 
\begin{align} 
\frac{\varphi}{f_\varphi} = \sqrt{ \frac{\sigma_u^2 + \sigma_d^2 + \sigma_s^2}{\sigma_{u0}^2 + \sigma_{d0}^2 + \sigma_{s0}^2}} 
=
\sqrt{ \frac{\sigma_u^2 + \sigma_d^2 + \sigma_s^2}{2 \sigma_{0}^2  + \sigma_{s0}^2}} 
\,, 
\end{align} 
where $\sigma_{f0}$ denotes the vacuum expectation value of $\sigma_f$ at 
$T=eB=0$ with $\sigma_{u0}=\sigma_{d0} = \sigma_0$, 
while $\sigma_f$s are the background fields depending on $T$ and $eB$.

$T_\mu^\mu(M_f, T, eB)$ in Eq.(\ref{EM-scale-ano}) is the trace anomaly 
induced from the quark loop with the constituent mass $M_f$ in Eq.(\ref{Mfs}) 
in the thermomagnetic medium with a weak $eB \ll (4 \pi f_\pi)^2$, which takes the form~\cite{Kawaguchi:2021nsa}
\begin{align} 
T_\mu^\mu(M_f, T, eB) 
= 
\widetilde{\beta}_e \cdot |eB|^{2}+\frac{1}{2} \sum_{f} q_{f}^{2} F\left(T, M_{f}\right)|e B|,\label{Tmumu}
\end{align} 
where 
\begin{align}
\sum_{f} q_{f}^{2} F\left(T, M_{f}\right) 
&=\sum_{f} q_{f}^{2} \frac{4 N_{c}}{\pi^{2}}\left[\left(M_{f} T \tilde{a}\left(M_{f} / T\right)-T^{2} \tilde{c}\left(M_{f} / T\right)\right)+\frac{1}{4} M_{f}^{2} \tilde{b}\left(M_{f} / T\right)\right], \notag \\
\tilde{a}\left(M_{f} / T\right) &=\ln \left(1+e^{-M_{f} / T}\right), \notag \\
\tilde{b}\left(M_{f} / T\right) &=\sum_{r=1}^{\infty}(-1)^{r}\left(-\int_{r M_{f} / T}^{\infty} \frac{e^{-t}}{t} d t\right), \notag \\
\tilde{c}\left(M_{f} / T\right) &=\sum_{r=1}^{\infty}(-1)^{r} \frac{e^{-r M_{f} / T}}{r^{2}}. \notag  
\end{align} 
In Eq.(\ref{Tmumu}) $\widetilde{\beta}_e$ in the first term of Eq.(\ref{Tmumu}) stands for  
 a beta-function coefficient for the electromagnetic coupling $e$ divided by $e^3$. 
When evaluated at the one-loop level, $\widetilde{\beta}_e$ is independent of the electromagnetic coupling $e$:  
\begin{align} 
\widetilde{\beta}_e &=\frac{1}{(4\pi)^2}\frac{4N_{c}}{3} \sum_{f}q_{f}^{2}
\,. \label{betatilde}
\end{align} 
The first term in Eq.(\ref{Tmumu}) is nothing but the well-known 
electromagnetic scale anomaly at the magnetized vacuum with a constant magnetic field $eB$, 
while the second term corresponds to the thermomagnetically induced part~\cite{Kawaguchi:2021nsa}. 
The function $F(T, M_f)$ is positive definite and monotonically increases with $T$, 
hence the tadpole term in Eq.(\ref{EM-scale-ano}) keeps developing with respect to both $T$ and $eB$.

Both of two electromagnetic-scale anomaly terms 
originally take the form of the transverse polarization like $F_{\mu\nu}^2$ in the operator form, reflecting  
the electromagnetic gauge invariance. 
Therefore, they will tend to vanish as the magnetic field 
gets strong enough, because the dynamics of quarks are governed by 
the lowest Landau level states longitudinally polarized along the direction 
parallel to the magnetic field. 
Thus, the electromagnetic scale anomaly is 
most prominent in a weak magnetic regime, as noted in the literature~\cite{Kawaguchi:2021nsa}.

Before proceeding, we shall give several comments on the validity of 
the present formulation as the thermomagnetic chiral model in light of 
properly investigating the chiral phase transition around 
the three-flavor chiral limit. 

\begin{itemize} 

\item

In Eq.(\ref{EM-scale-ano}) 
the lightest-chiral singlet-$\varphi$ meson has been assumed to be mostly 
composed of the quarkonium state ($\bar{q}_fq_f$).  
A glueball field could mix with $\varphi$, 
hence might contribute to the scale anomaly 
relevant to the chiral phase transition. 
However, lattice simulations for 
$2 + 1$ flavors have clarified that 
around the two-flavor chiral limit with $m_0$ less than 
the physical-point value, 
the fluctuation of Polyakov loop, which can be regarded 
as (an electric part) of glueball, does not have significant correlation with   
the chiral order parameter~\cite{Clarke:2019tzf}. 
This indicates negligible mixing between the glueball  
and $\varphi$ in the two-flavor chiral limit that in the present work 
we focus mainly on.   
Hence such a gluonic term will not be taken into account later on.

\item 

Though the mixing with a glueball is irrelevant, one might still suspect 
a direct contribution from the gluonic scale-anomaly (of the form $\sim \varphi G_{\mu \nu}^2)$ to the chiral phase transition, 
which is thought to be present all the way in thermal QCD including the 
vacuum. 
It is known that the gluonic scale-anomaly can be introduced as the log potential form in a scale-chiral Lagrangian~\cite{Brown:1991kk}. 
The log potential form like $\varphi^4 \log \varphi $ 
does not give any interference for 
the electromagnetic scale-anomaly induced tadpole-term which dominates around $\varphi \sim 0$.

\item 
In deriving the thermally induced tadpole part in Eq.(\ref{Tmumu}), we have ignored 
the next-to-leading order terms of  ${\cal O}((eB)^2)$~\cite{Kawaguchi:2021nsa}, which might come in the formfactor $F$ as the nonlinear $eB$ dependence. 
Such higher order corrections would naively 
be suppressed by a loop factor, compared to the leading order term of 
${\cal O}(eB)$ in Eq.(\ref{Tmumu}). 
Even if it would not sufficiently be suppressed, and 
would constructively enhance or destructively weaken the tadpole contribution, 
the presently addressed chiral-phase transition nature 
would not substantially be altered, as long as the magnetic field 
is weak enough. 
To be more rigorous, it is subject to nonperturbative analysis on  
the photon polarization function coupled to the sigma field. 
In Ref.~~\cite{Ghosh:2019kmf}
the authors have made an attempt to straightforwardly compute the photon polarization in the weak field approximation at finite temperature. 
It would be interesting to apply such a technique described 
in the literature to our study, which is, however, beyond the scope of the present paper, instead deserve to another work. 

\end{itemize}

\subsubsection{Total form of thermomagnetic potential} 

The thermomagnetic potential containing the tadpole term in Eq.(\ref{EM-scale-ano}) in the mean field approximation thus takes the form  
\begin{equation}
\begin{aligned}
\Omega\left(\phi_u, \phi_d, \phi_s, T, e B\right)=&\sum_{f=u, d, s} \Omega_{f}\left(\phi_{f}, T, e B\right) +2 G\left(\phi_{u}^{2}+\phi_{d}^{2}+\phi_{s}^{2}\right)\\&-4 K \phi_{u} \phi_{d} \phi_{s}+V_{\mathrm{eff}}^{(\mathrm{Tad})}\left(M_{f}, T, e B\right),
\end{aligned}
\label{Omega-total}
\end{equation}
where
\begin{equation}
\begin{aligned}
\Omega_{f}\left(\phi_{f}, T, e B\right)= 
- N_{c} \frac{\left|q_{f} e B\right|}{2 \pi} \sum_{n=0}^{\infty} \alpha_{n} \int \frac{\mathrm{d} p_{z}}{2 \pi} 
\\ 
\times\left[E_{q_{f}, n}+2 T \ln \left(1+\exp \left(-\frac{E_{q_{f}, n}}{T}\right)\right)\right]
\,,
\end{aligned}
\end{equation} 
with $E_{q_{f}, n}^{2}=p_{z}^{2}+2 n\left|q_{f} e B\right|+M_{f}^{2}$ being the energy dispersion relation for the $f$-quark, and 
$\alpha_{n}=2-\delta_{n, 0}$, which counts the spin degeneracy with respect to 
Landau levels. 
The thermomagnetic potential is thus evaluated as a function of quark condensates 
$\phi_u, \phi_d$, and $\phi_s$ through Eq.(\ref{Mfs}) with Eq.(\ref{Sf}). 
Given $m_s$, $T$, and $eB$, we search for the minimum of the thermomagnetic potential in Eq.(\ref{Omega-total}) by focusing on the chiral limit 
for up and down quarks, $m_0 \to 0$, through the stationary condition, 
\begin{equation}
\frac{\partial \Omega\left(\phi_{u}, \phi_{d}, \phi_{s}, T, e B\right)}{\partial \phi_{f}} 
\Bigg|_{m_0 =0} =0 
\,. 
\end{equation}

\section{Disappearance of the first-order chiral phase transition}

In this section, we show that the first-order feature 
of the chiral phase transition in the light or massless three-flavor case 
is wiped out in the presence of a weak magnetic field background. 
We start with a heuristic approach to qualitatively 
understand the essential difference on the type of the phase transition orders between the cases with and without 
a weak background magnetic field in the massless three-flavor limit [See Fig.~\ref{schematic}]. 
Then we confirm our observation explicitly by numerical analysis [See Fig.~\ref{Mu-T}], 
and extend the parameter space by including finite strange quark mass $m_s$, which is projected 
onto an extended Columbia plot in the ($eB/f_\pi^2$, $m_s/f_\pi$) plane 
[See Fig.~\ref{11}].

\subsection{Heuristic observation: Overview of the disappearance}
\label{Heuristic}

\begin{figure}[t] 
\centering 
\includegraphics[width=0.48\textwidth]{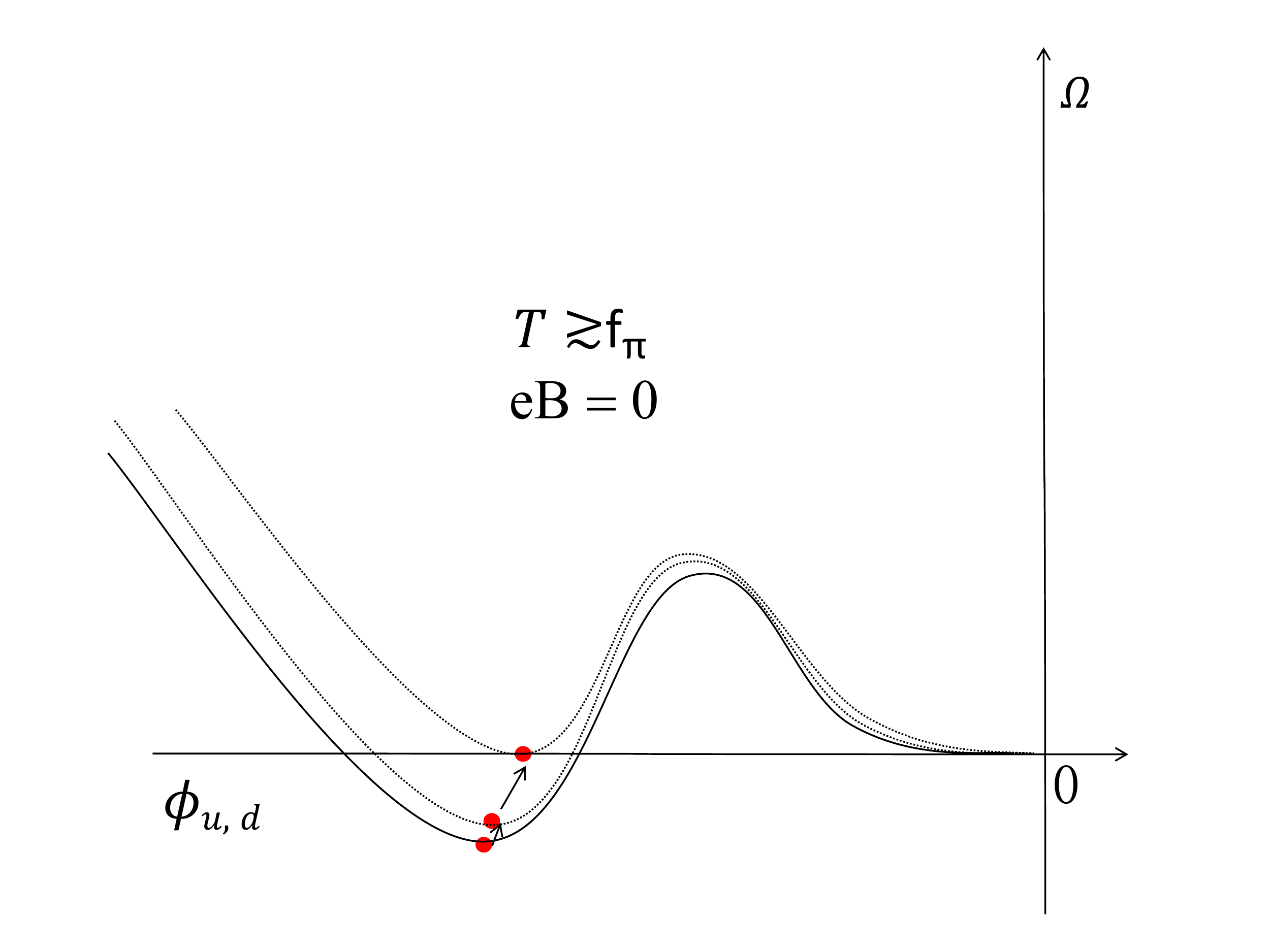}
\includegraphics[width=0.48\textwidth]{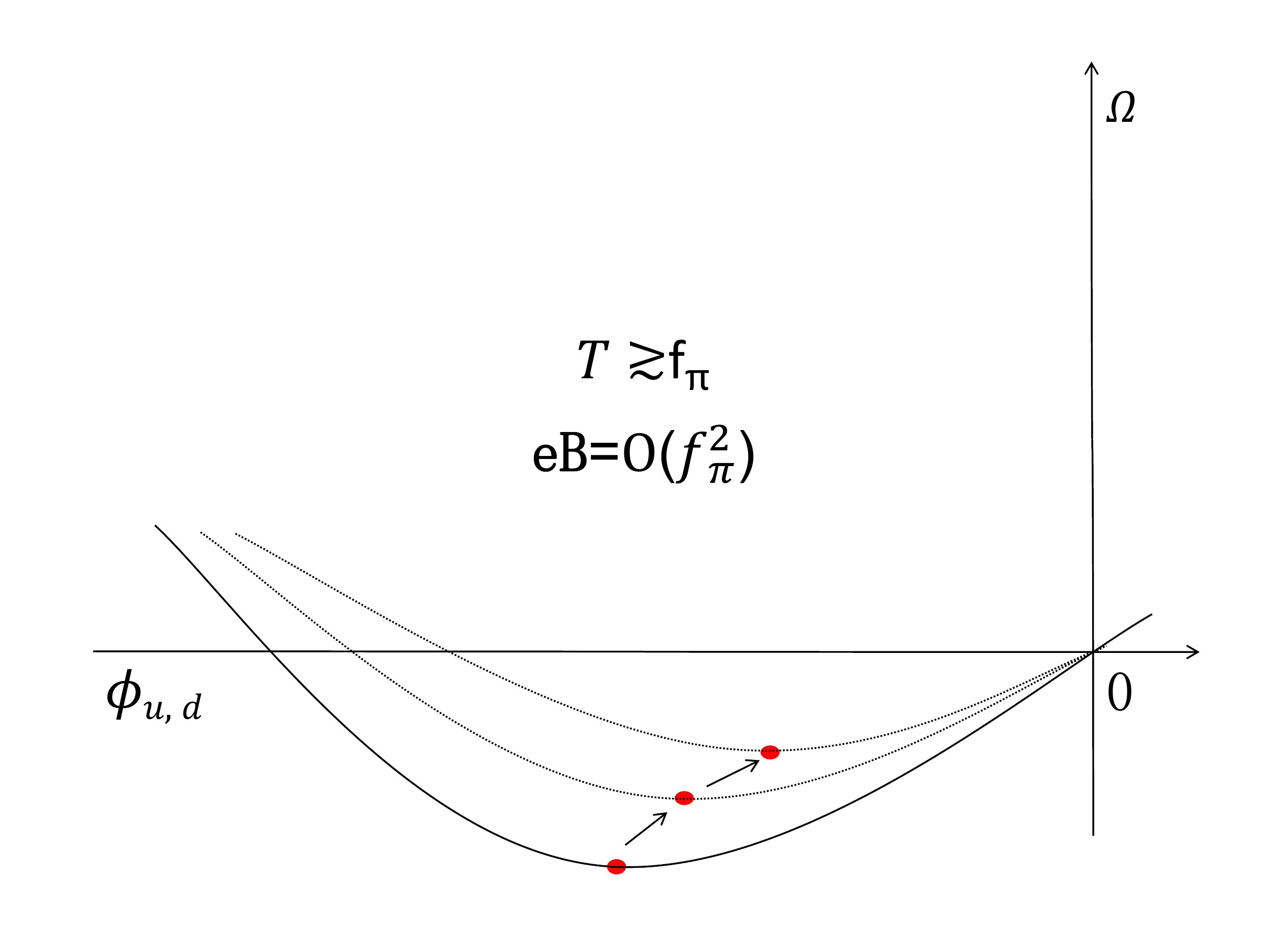}
\caption{An illustration of schematic view of the potential deformation 
around the chiral phase transition at $T \gtrsim f_\pi$ 
in the massless three-flavor QCD, 
with $eB =0$ (left panel) and $eB= {\cal O}(f_\pi^2)$. 
The thermodynamic (magnetic) potential $\Omega$ (the vertical axis) is described 
as a function of the chiral order parameters $\phi_u$ and $\phi_d$ (the horizontal axis). The arrows in the panels depict the evolution of the vacuum 
when $T$ gets higher and higher. 
In the left panel with $eB=0$, the first-order transition is realized due to the 
sizable cubic potential arising from the KMT term, while in 
the right panel with $eB= {\cal O}(f_\pi^2)$ 
the transition becomes smoother, like the second order or crossover, 
because the cubic term fails to keep significant against the magnetically  
enhanced quadratic term which destructively contributes over the thermal correction 
(due to the magnetic catalysis), and only the single vacuum is favored due to the presence of the 
`magnetic-scale anomaly-induced tadpole term. 
For more details, see the text.  
} 
\label{schematic}
\end{figure}

First of all, consider massless three-flavor QCD with $eB=0$, and note the presence of the cubic potential ($\sigma^3$) induced from 
KMT determinant term, where $\sigma$ stands for a symbolic order parameter regarding the chiral $SU(2)_L \times SU(2)_R$ symmetry,  
such as up and down quark condensates $\phi_u$ and $\phi_d$ in Eq.(\ref{Mfs}) 
as in the Ginzburg-Landau description. 
This cubic term will not get any correction from the thermal-quark loops, 
hence keeps contributing to the potential at any $T$ with negative sign like 
$V_{\sigma^3} = - |c| \cdot \sigma^3 $ (with some coefficient $|c|$) 
which is definitely fixed by the positive sign of the $\eta'$ mass squared. 
This is because the thermal quark loop correction does not break the chiral $SU(2)_L \times SU(2)_R$ symmetry or the $U(1)_{A}$ axial symmetry in the mean field approximation. 
In contrast, 
the quadratic $(\sigma^2)$ and quartic ($\sigma^4$) terms get 
sizable thermal corrections, which make the potential lifted up and the potential energy at the nontrivial vacuum smaller in magnitude as $T$ gets higher. 
Thus the thermal corrections help keeping the potential barrier created by the cubic term sizable enough to stay even at 
high $T$, such that the size of the $\sigma^3$ term becomes comparable to those of the $\sigma^2$ and $\sigma^4$ terms in the potential, and then the phase transition happens at the first order with 
a jump from $\sigma \neq 0$ to $\sigma=0$. 
This feature at around the critical temperature is generically seen irrespective to the sign of the original quadratic term at $T=0$ 
and the size of the cubic term. 
See the left panel of Fig.~\ref{schematic}, which would help qualitatively grasping  the potential deformation around the phase transition era.

Next, turn on a weak $eB$, say, as weak as the strength of ${\cal O}(f_\pi^2)$.  
Two remarkable natures, which act destructively against the thermal corrections,  will then be seen: 

\begin{itemize} 

\item [i)] 
One is the magnetic catalysis, which makes the chiral-broken nontrivial vacuum and the vacuum-potential well deeper as $eB$ increases against the thermal corrections (i.e. makes the chiral breaking promoted)~\footnote{
Lattice QCD has observed the inverse magnetic catalysis which acts constructively with increase of $T$ at high enough $T = {\cal O}(100\,{\rm MeV})$~\cite{Bali:2011qj,Bornyakov:2013eya, Bali:2014kia, Tomiya:2019nym, DElia:2018xwo, Endrodi:2019zrl}. 
The magnetic field strength is bounded from below as $eB \gtrsim T^2$ due to the currently 
applied method to create the magnetic field~\cite{Endrodi}, 
which is only for the physical pion mass simulations at finite temperatures. 
Thus applying a weak $eB$ at high $T$ with much smaller pion masses, particularly close to the chiral limit, is challenging due to the higher numerical cost, 
and it is still uncertain whether the normal or inverse magnetic 
catalysis persists. 
}; 

\item[ii)] 
The other is emergence of the thermomagnetic tadpole in Eq.(\ref{EM-scale-ano}), 
which makes the chiral-symmetric vacuum $\sigma =0$ no longer the vacuum by the 
negative linear slope at $\sigma=0$ [See Eq.(\ref{Tmumu})]. 

\end{itemize} 
Due to the phenomenon i), the $\sigma^2$ term in the potential gets negatively and significantly enhanced. Convoluted with the second feature ii), as $eB$ increases, 
the $\sigma^3$ term will then lose the chance to become comparable with the $\sigma^2$ term and fail to jump into another vacuum even at high $T$, thus 
there is only the single vacuum, and 
the phase transition will be smoother like the type of crossover or 
the second order. See the right panel of Fig.~\ref{schematic}~\footnote{
\textcolor{black}{
The magnetic catalysis in i) would also play an crucial role even when 
the gluonic scale anomaly contribution is incorporated in the system 
following the procedure as in a chiral-scale effective theory~\cite{Brown:1991kk}: the thermal mass term $\sim T^2 \varphi^2$   
gets destructively interfered with a (relatively) strong $eB$, enough not to build a potential barrier which could be created when the logarithmic 
potential term $\sim \varphi^4 \log \varphi$ associated with the 
gluonic scale anomaly is present. 
}
}.

This is the overview of what will happen in the presence of 
a weak magnetic field along with the induced tadpole potential, 
in conjunction with the magnetic catalysis, which is robust and generic, and thus 
{\it the first-order nature of the chiral phase transition in 
the massless three-flavor QCD disappears}.

\subsection{Analysis in the massless three-flavor limit}	
\label{Sec3.2}
  
Using the QCD-monitor parameter set in Eqs.(\ref{parameters1}) and (\ref{parameters2}),  
we evaluate the $T$ dependence 
of the constituent mass of up quark $M_u$ in Eq.(\ref{Mfs}) 
with $eB$ varied. 
Note that the chiral symmetry is explicitly broken even in the massless limit, 
because of the magnetic field, hence the chiral restoration cannot exactly be achieved even at high enough $T$ and the criticality becomes pseudo. 
We therefore determine the order of the phase transition 
by reading off the sharpness of $d M_u/dT$ at the pseudocritical temperature $T_{\rm pc}$ defined 
at the inflection point as $d^2 M_u/dT^2|_{T=T_{\rm  pc}} =0$.  
When a discontinuous jump in $d M_u/dT|_{T=T_{\rm pc}}$ is observed, 
we interpret the phase transition to be of first order, 
otherwise a smooth transition including the second order 
and crossover.

The analysis has been done fully in terms of dimensionless quantities normalized to the pion decay constant at the vacuum ($T=eB=0$), $f_\pi$: all relevant things are scaled by the powers of $f_\pi$ and can be evaluated as a function of  
$\frac{T}{f_\pi}, \frac{eB}{f_\pi^2}, \frac{m_s}{f_\pi}, \frac{M_f}{f_\pi}$, and $\frac{\phi_{f}}{f_\pi^3}$ 
together with the other dimensionless model parameters in Eqs.(\ref{parameters1}) and (\ref{parameters2}). 

\textcolor{black}{
 Note that the electromagnetic coupling $e$ itself is a redundant parameter, 
because it always arises with $B$ as the combination of $eB$ as long as the gauge interaction 
is perturbatively small enough to be well approximated at the one-loop level as 
in Eq.(\ref{betatilde})}.

Thus our results will be fairly generic 
and can be applied to a wide class of QCD-like theories including 
dark QCD coupled to dark photon with charges ($q_f$) of ${\cal O}(1)$ like 
the electromagnetic charges of the ordinary quarks.

\begin{figure}[t]
\centering 
\includegraphics[width=0.49\textwidth]{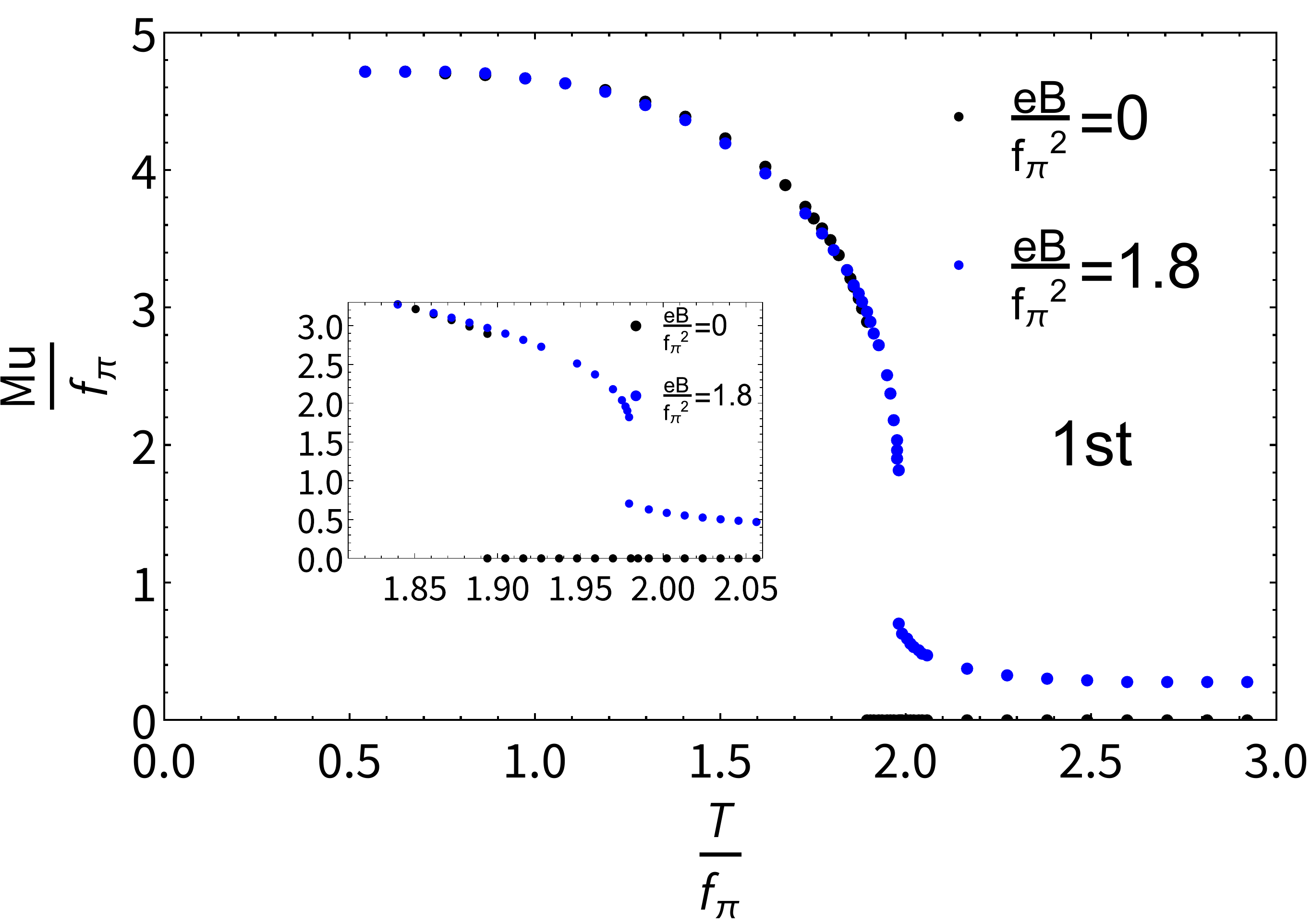}
\includegraphics[width=0.49\textwidth]{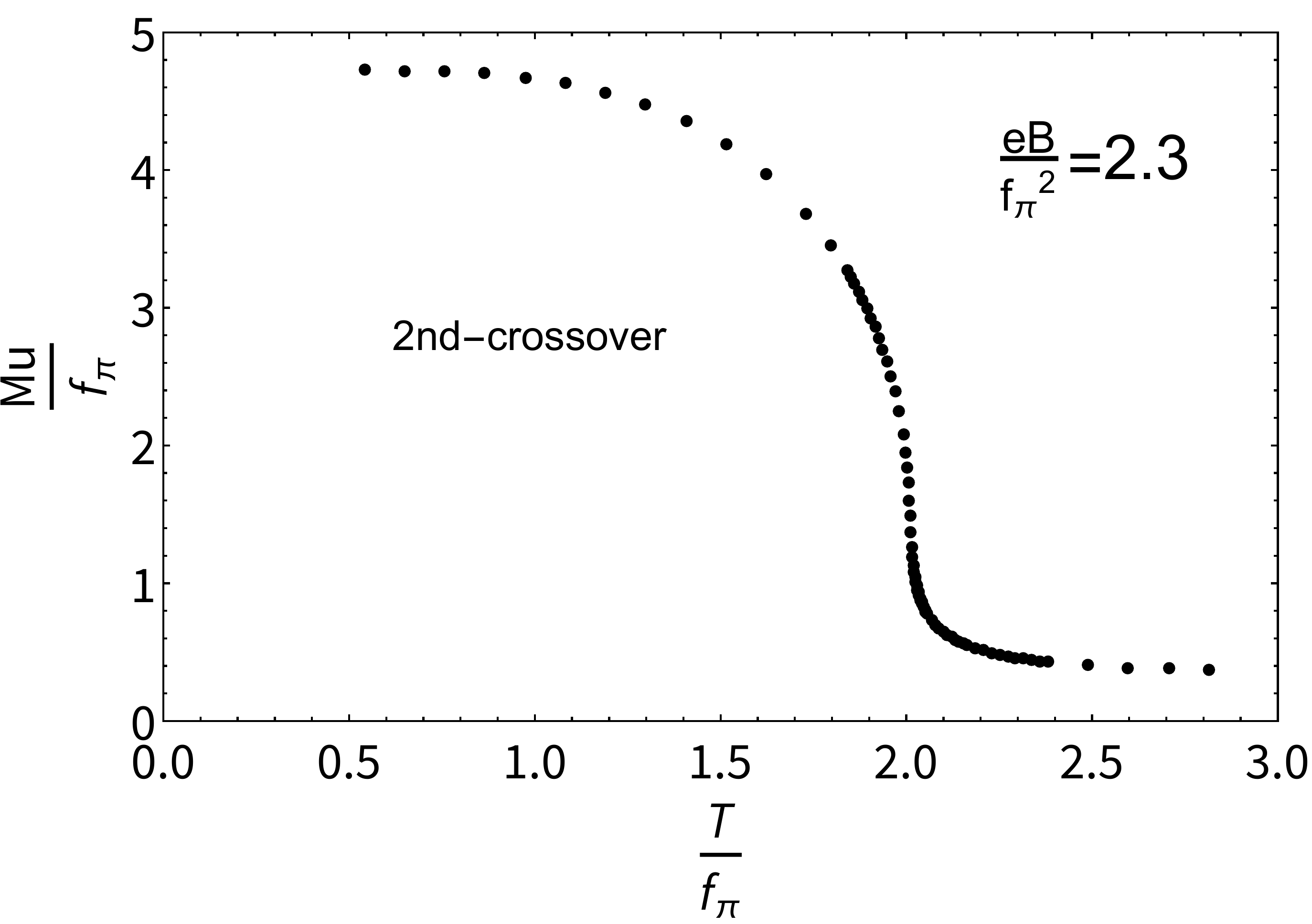}
\caption{ The $T$ dependence of $M_u$, normalized by the pion decay constant at the vacuum with $T=eB=0$, $f_\pi$, 
in the case with massless three flavors (with $m_0=m_s=0$) for 
$eB/f_\pi^2 = 1.8$ (blue curve), in comparison with the case with $eB=0$ (black curve)  
(left panel: the first order transition), and $eB/f_\pi^2 = 2.3$  
(right panel: the second order transition or crossover). 
The close-up window in the left panel zooms-in the difference 
between the cases with $eB=0$ and $eB/f_\pi^2= 1.8$ around the 
critical temperature. 
}
\label{Mu-T} 
\end{figure}

In Fig.~\ref{Mu-T} we show the $T$ dependence of $M_u$, normalized to $f_\pi$, 
in the massless three-flavor limit ($m_0=m_s=0$), 
for a couple of reference cases with $eB/f_\pi^2 = 1.8$ (left panel) and 2.3 (right panel). 
We have confirmed that numerical results are unchanged even raising  
the Landau level up to $n = 500$ for $eB/f_\pi^2 < 1.8$ including 
$eB=0$, which means that our calculations 
are accurate enough. 
The first-order phase transition can still be seen in the left panel, 
however, in the right panel with the larger $eB$ the transition gets changed to 
be smoother, which is of the type of the second order or crossover, 
in accordance with the heuristic observation in Sec.~\ref{Heuristic} [See Fig.~\ref{schematic}].

The strength of the first-order phase transition 
can be evaluated by measuring the size of 
the ratio $\Delta M_u/T_{\rm pc}$ with $\Delta M_u = M_u^{\rm before}(T_{\rm pc}) 
- M_u^{\rm after}(T_{\rm pc})$ 
being the difference 
of the degenerated values of $M_u$ before and after the phase transition at $T=T_{\rm pc}$. 
From the left panel of Fig.~\ref{Mu-T}, 
in the case with $eB/f_\pi^2 = 1.8$ (blue curve)
compared to the case with $eB=0$ (black curve), 
it reads 
\begin{align} 
\frac{   \Delta M_u/T_{\rm pc} |_{ eB = 1.8 f_\pi^2 }}{
\Delta M_u/T_{\rm pc} |_{eB =0}
} 
\simeq \frac{0.56}{1.5}
\simeq  0.37
\,. \label{latent-heat}
\end{align} 
This implies that the strength of the first-order phase transition gets reduced 
to be about 37\% of the one at $eB=0$, 
where the potential is normalized as $\Omega[M_u=M_u^{\rm after} (T_{\rm pc})] = 0$.  This happens due to two phenomena:  
i) the chiral symmetric vacuum ($M_u=0$) is no longer realized 
in a magnetic field even at high $T$;  
ii) the temperature required to make $M_u$ jumped down gets higher to further 
decrease $M_u$ until the transition takes place, which is  
due to the magnetic catalysis, so that $\Delta M_u$ becomes smaller.  
Thus the strong first-order feature in massless-three flavor QCD 
(with $\Delta M_u/T_{\rm pc} > 1$) significantly gets altered by increasing a weak $eB$ to be finally gone as in 
the right panel of Fig.~\ref{Mu-T}.

The reduction of the strength of the first-order phase transition 
further leads to  weakening of the strength of the produced latent heat $L(T)$, which is evaluated  
with fixed $eB/f_\pi^2$ as 
\begin{align} 
 \frac{L(T_{\rm pc})}{T_{\rm pc}^4} & 
 =  - \frac{\Delta \Omega (T_{\rm pc})}{T_{\rm pc}^4} + 
 \frac{1}{T_{\rm pc}^3} \frac{\partial \Delta \Omega (T)}{\partial T} \Bigg|_{T=T_{\rm pc}} 
 \,, \notag\\ 
{\rm with} \qquad \Delta \Omega(T) & \equiv  
\left[ \Omega (\phi_{u,d,s}^{\rm before}, T, eB) 
 - \Omega(\phi_{u,d,s}^{\rm after}, T, eB) \right]_{eB=0, \, {\rm or}\, 1.8 f_{\pi}^2} 
\,.
\end{align}
\textcolor{black}{Then we find 
\begin{align} 
 \frac{L(T_{\rm pc})/T_{\rm pc}^4|_{eB = 1.8 f_\pi^2}}{L(T_{\rm pc})/T_{\rm pc}^4|_{eB = 0}}    
\simeq 0.85 
\,. \label{latent-heat2}
\end{align}  
Thus the gravitational wave spectrum created via the bubble nucleation associated with this dark chiral phase transition gets weaker (by about 15\%)}. 
The associated generation mechanism for the dark baryon - anti-dark baryon asymmetry 
and the stable formation of dark baryonic compact stars and MACHOs could also be affected.

\subsection{Fate of the first-order regime in a view of Columbia plot}

\begin{figure}[t]
	\centering
	\includegraphics[width=0.55\textwidth]{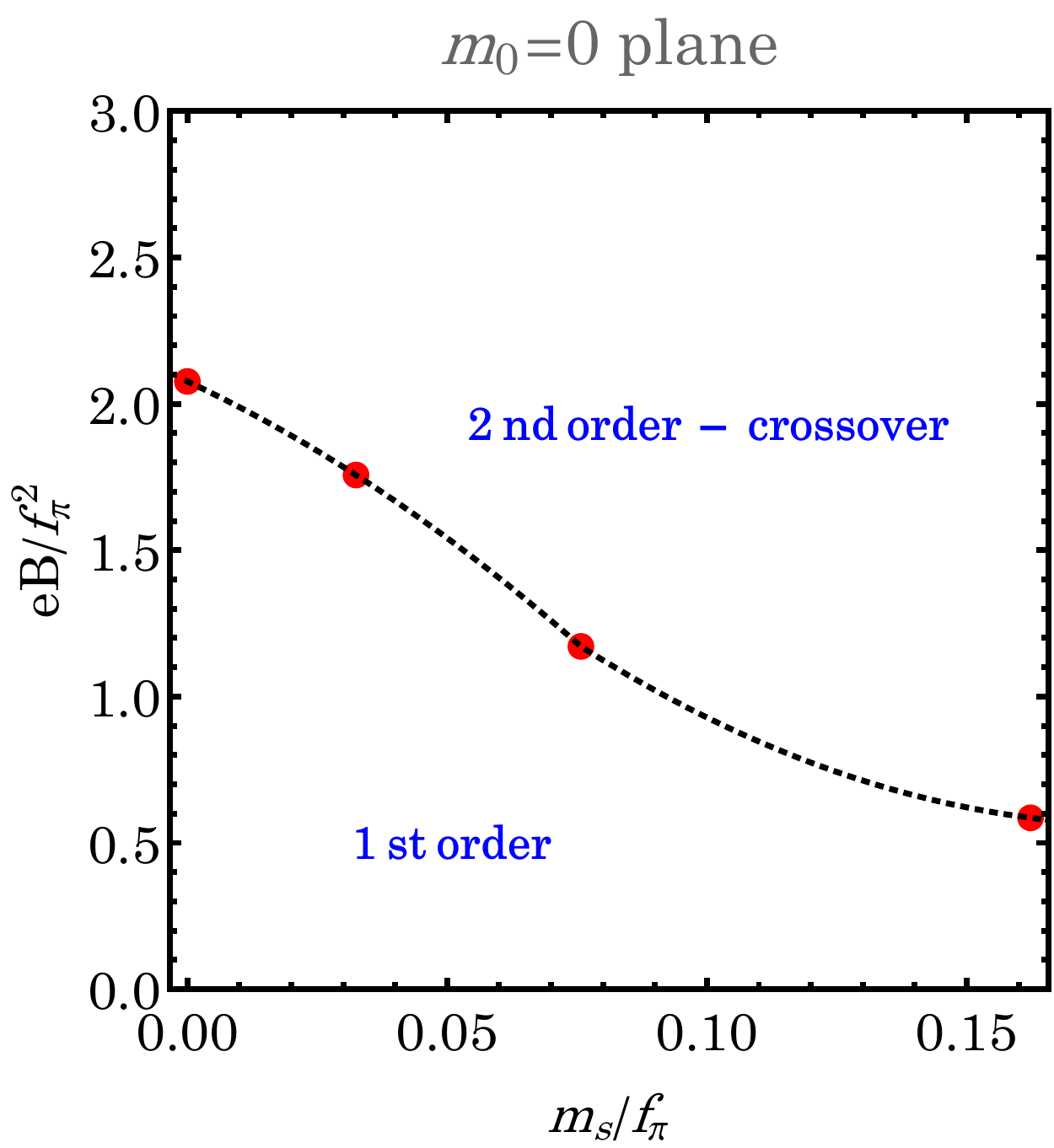}
	\caption{
	\textcolor{black}{
	The phase diagram in a view of an extended Columbia plot projected onto the $(m_s/f_\pi, eB/f_\pi^2)$ parameter space on the $m_0=0$ plane. 
	The four reference points (denoted as red blobs) have been chosen 
	at $(m_s/f_\pi, eB/f_\pi^2) \simeq$ 
(0.162, 0.586),  
(0.0758, 1.17),   
(0.0325, 1.76), and  
(0, 2.08),  
to make the interpolation curve for the phase boundary (corresponding to the black dots) which separates the first order and second order - crossover phases. 
} }
	\label{11}
\end{figure}

Further allowing $m_s$ to take finite values, in Fig.~\ref{11} we make an extended Columbia plot on the $(m_s/f_\pi, eB/f_\pi^2)$ plane in the two-flavor chiral limit with $m_0=0$. 
\textcolor{black}{
We have taken 
the representative four points 
$(m_s/f_\pi, eB/f_\pi^2) \simeq$ 
(0.162, 0.586),  
(0.0758, 1.17),  
(0.0325, 1.76), and (0, 2.08), 
and interpolated those reference points to draw the 
phase boundary separating the first order and second order - crossover regimes.}

\begin{figure}[t]
	\centering
	\includegraphics[width=0.55\textwidth]{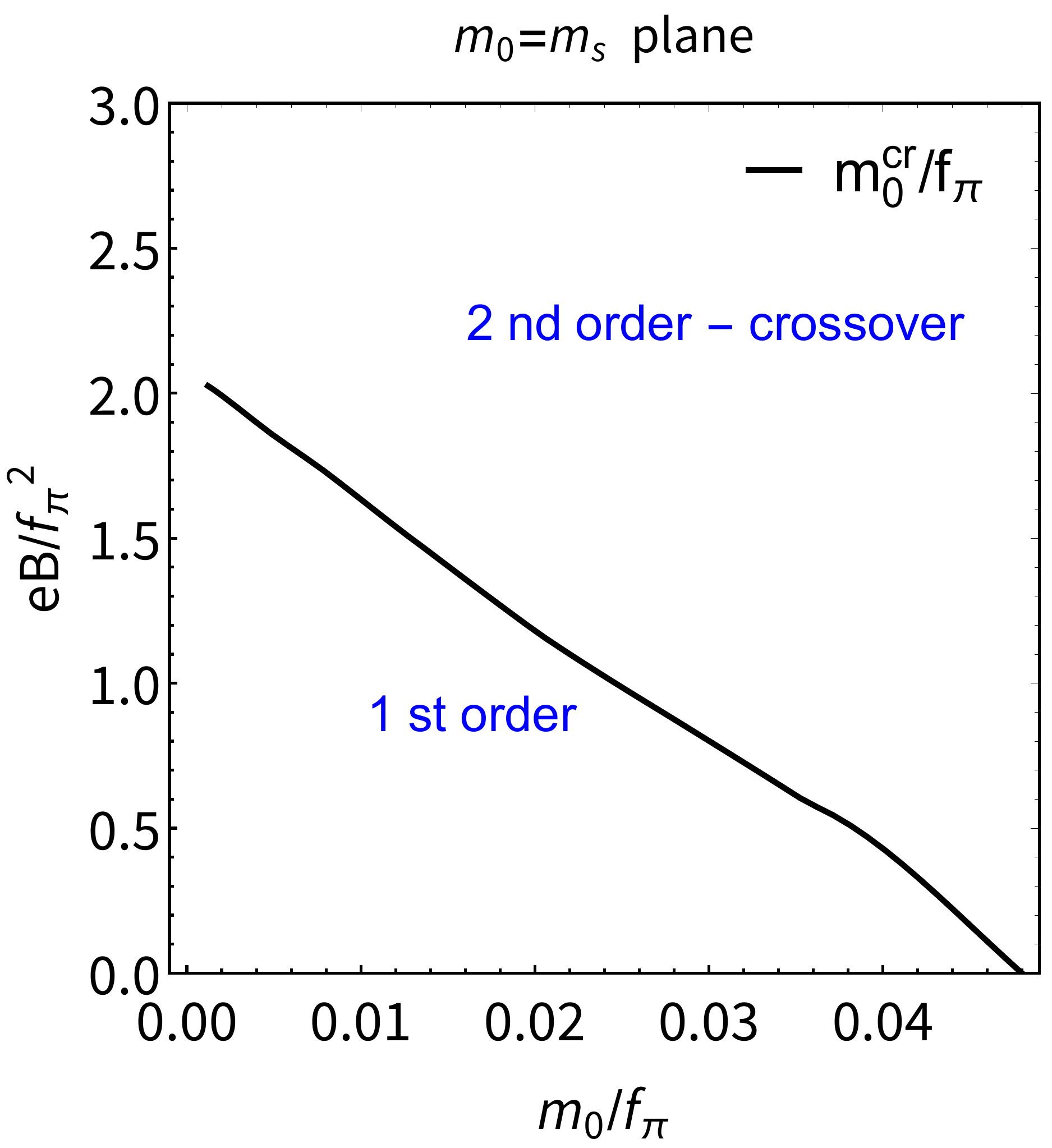}
	\caption{
	\textcolor{black}{Another look at the phase diagram in terms of an extended Columbia plot, on the $m_0=m_s$ plane (in the three-flavor symmetric limit)
	in the $(m_0/f_\pi, eB/f_\pi^2)$ space. The black curve corresponds to the critical surface separating the two phases, the first order and second order - crossover regimes. }
	}
	\label{13}
\end{figure}


\textcolor{black}{
The range of $m_s/f_\pi$ to realize 
the first-order phase transition gets narrower and narrower, 
as $eB$ increases. 
This trend confirms the conjecture made in  Ref.~\cite{Kawaguchi:2021nsa} based on the quark-meson model, 
where at $eB=f_\pi^2$ a new critical endpoint (line) has been conjectured 
to be present at a finite $m_s$ on the $m_s$ axis in the extended Columbia plot, 
which corresponds to the ``2nd order - crossover" domain 
in Fig.~\ref{11} including the massless two-flavor limit with extrapolation of $m_s \to \infty$}~\footnote{
We have found that the strange quark completely decouples from the system at $m_s/f_\pi \simeq 216$. }.

\textcolor{black}{
The first-order regime completely disappears
 when $eB$ reaches the critical strength, 
 $ eB_{\rm cr}/f_\pi^2 \simeq 2.1$.    
This probes the disappearance of 
the first-order regime due to the convolution of the magnetic catalysis 
with the electromagnetic scale anomaly, as has heuristically been viewed 
in Sec.~\ref{Heuristic}.  
}

\textcolor{black}{
We have also examined the critical quark mass 
in the three-flavor symmetric limit ($m_0=m_s$), at which the phase transition feature  turns from the first order to the second-order, to be crossover. 
In the case with $eB=0$ we have 
$m_{0}^{\rm cr}|_{eB=0}/f_\pi \simeq 0.048 
$. 
Applying a weak $eB$, we find that $m_0^{\rm cr}$ monotonically 
decreases as $eB$ increases as plotted in Fig.~\ref{13}. 
This is consistent with Fig.~\ref{11}, which shows that the first-order domain (the range of $m_s/f_\pi$) gets shrunk as 
the strength of the magnetic field gets stronger. 
Indeed we have observed $m_0^{\rm cr}$ at the critical magnetic field $eB_{\rm cr}/f_\pi^2 \simeq 2.1$. 
Increasing the size of the quark mass allows even weaker $eB$ 
to cease the phase transition at the first order, 
as seen in the phase diagram, Fig.~\ref{13}. 
}

Figures~\ref{11} and \ref{13} thus imply that 
the critical magnetic field $eB_{\rm cr}/f_\pi^2 \simeq 2.1$ brings 
complete disappearance of the first-order nature in the small quark mass 
regime present in the conventional Columbia plot 
without a magnetic field~\cite{Columbia}.



\section{Conclusion and outlook}
In conclusion, 
thermal QCD with light three flavors may not undergo the first-order chiral phase transition,  
when ordinary or dark quarks are externally coupled to a weak enough background field strength of photon or dark photon, which we collectively call a background ``magnetic" field.  
The critical ``magnetic" field strength $(eB_{\rm cr})$ normalized to the pion decay constant at the  vacuum with $T=eB=0$, $f_\pi$, is found: $eB_{\rm cr}/f_\pi^2 \sim 2$, above  
which (but at still weak $eB$ $\ll (4 \pi f_\pi)^2$) the first-order feature goes away [See Figs.~\ref{11}, \ref{13}, and also Eqs.(\ref{latent-heat}) and (\ref{latent-heat2})]. 
\textcolor{black}{
As quarks get massive, weaker $eB$ enables to wipe out the first-order feature [See Fig.~\ref{13}]. 
}

The disappearance of the first-order regime 
is the generic consequence of the crucial presence 
of the ``magnetically" -induced scale anomaly and the ``magnetic" catalysis for the chiral symmetry breaking at around the criticality, 
which is irrespective to details of the parameter setting 
such as those in Eqs.(\ref{parameters1}) and (\ref{parameters2})
[See also Fig.~\ref{schematic}]. 
Indeed, the trend of the disappearance of the first-order feature 
as shown in Fig.~\ref{11} (or \ref{Mu-T}) is fairly generic  
and can be applied to a wide class of QCD-like theories,  
\textcolor{black}{which might still hold even beyond the mean-field approximation in the NJL description}.

\textcolor{black}{
The dark photon is typically the portal between the dark and the Standard Model sectors. 
In particular, the dark photon coupling $e$ is crucial to drain the entropy of the Standard Model particles in the 
Universe. 
This generically constrains the value of $e$ for dark QCD models such as SIMP models to be phenomenologically and cosmologically viable. 
As noted in Sec.~\ref{Sec3.2}, 
the present analysis is irrespective to 
the size of $e$ itself, because the external-magnetic field strength 
contributes to dark QCD models with keeping the gauge-invariant form $F_{12} = eB$ without separation between $e$ and $B$. 
Therefore, the presence of the critical strength $eB_{\rm cr}/f_\pi^2$ and the extended Columbia plots in Figs.~\ref{11} and~\ref{13} do not give any constraint on a singled $e$, or a dark photon portal, hence 
are generic consequences applicable for any size of dark photon coupling $e$ in any 
dark QCD scenario. 
} 

\textcolor{black}{
When the ``magnetogenesis" is realized by some inflationary scenario along with reheating Universe, the initially produced strength of $\sqrt{eB}_{0}$ could be on the order of the reheating temperature. Passing through the redshift down to the dark QCD epoch, one then considers the critical condition $(eB_0 \gg) eB_{\rm cr} \sim 2 f_\pi^2$. 
This implies that our finding can 
constrain the magnetogenesis and the reheating scenario, instead of the dark photon physics in dark QCD.}

The present conclusion would thus give significant impacts on modeling and testing dark QCD via the astrophysical probes and/or constraining the magnetogenesis, as well as would  help deeply understanding the chiral phase structure in QCD.

In closing, we give several comments on the future prospects in order. 

\begin{itemize}

\item 
The chiral extrapolation on lattice 
in absence of a magnetic field 
has systematically been studied, which has also been improved so much  
in the case with $2+1$ flavors \cite{Sheng-Tai}. 
In addition, a remarkable progress on creating small magnetic 
fields at the physical point has been reported, 
and the current lower bound is 
$\sim$ 100 MeV \cite{Endrodi}. 
Calculations on derivatives with respect to magnetic fields 
have also been worked out directly on the lattice, 
so in principle small magnetic fields become accessible~\cite{Bali:2020bcn}. 
Therefore, we expect that 
the disappearance of the first-order nature could be explored 
on lattice QCD in near future.

\item 
The disappearance of the first-order feature 
also indicates null tricritical point at which 
all the first, second, and crossover regimes merge~\cite{Columbia}, 
so that the rich phase structure in the conventional Columbia plot  
is completely gone, and only the crossover nature survives.  
This phenomenon has been conjectured in the literature~\cite{Kawaguchi:2021nsa}, 
in which it is dubbed like {\it scale-anomaly pollution}. 
Currently this pollution gets more quantified in a sense that 
it has been found to take place at $eB$ above the critical strength, $eB \gtrsim 2 f_\pi^2$. 
It would be intriguing to examine how this pollution will go away as 
$eB$ increases up to the strength as much as the strong field regime ($\sim (4 \pi f_\pi)^2$), and the crossover feature can keep persisting. 
To this end, it would be necessary to formulate the electromagnetic-scale 
anomaly term as a formfactor depending on the photon-transfer momentum 
and polarization tensor structure, so that the decoupling of 
the Landau levels in the photon polarization function is made 
automatically embedded to interpolate the weak and strong magnetic field 
regimes continuously. 
Presumably, we then need to nonperturbatively compute the $\varphi$ - photon - photon
triangle diagram involving the nonperturbative quark propagators, which has never been carried out. 
Our present result would motivate people to work on this issue to 
clarify the fate of the scale-anomaly pollution as increasing $eB$ 
\textcolor{black}{
and also existence of a new tricritical point on the $eB$ -axis, 
as can be expected from Figs.~\ref{11} and~\ref{13}}.

\item
The current analysis has been done based on the NJL model 
in the mean field approximation. 
This approach is on the same level of reliability as in the literature~\cite{Holthausen:2013ota,Ametani:2015jla,Aoki:2017aws,Helmboldt:2019pan,Reichert:2021cvs,Aoki:2019mlt} in a sense of employing low-energy effective models to address the chiral 
phase transition instead of the underlying thermal dark QCD. 
Since the linear sigma model belongs to the same universality class 
as the NJL model in terms of the renormalization group, 
the current conclusion on the disappearance of the first-order nature 
could directly be applied also to the other dark QCD scenario
~\cite{Heikinheimo:2018esa,Bai:2018dxf,Archer-Smith:2019gzq,Dvali:2019ewm,Easa:2022vcw,Helmboldt:2019pan,Reichert:2021cvs,Tsumura:2017knk,Croon:2019iuh} argued based on the linear-sigma model description including 
the pioneer work by Pisarski and Wilczek~\cite{Pisarski-Wilczek}. 
Though being effective enough to take the first step, 
the current model analysis should anyhow be improved to be more rigorous 
by going beyond the mean field approximation, say, the nonperturbative renormalization group method. 
This way we would get a more conclusive answer to whether the first-order nature of 
the chiral phase transition surely goes away in 
thermomagnetic QCD with massless three flavors.

\item 
When contributions beyond the mean field approximation, or equivalently 
the subleading corrections in the $1/N_c$ expansion are taken into account, 
the meson loops would come into play in the chiral phase transition. 
If one then considers dark QCD with many flavors, 
the elimination of the first-order feature by a background ``magnetic" 
field could be avoided because a sizable thermal-cubic term amplified by the large number 
of flavors would be generated to build a strong enough potential barrier at around the critical temperature. 
This implies that the lower bound on the number of flavors might be placed 
with $eB/f_\pi^2$ fixed 
in such a way that 
a strong enough first order (with $\Delta M_u/T_{\rm pc} \gtrsim 1$) 
is realized.  
This would be interesting to study and give further impact on astrophysical 
and cosmological probes from dark QCD 
with many flavors~\cite{Bai:2018dxf,Archer-Smith:2019gzq,Dvali:2019ewm,Easa:2022vcw}. 
The issue can be simplified by working in the linear sigma model description, 
as an application of studies on many-flavor QCD in the literature~\cite{Kikukawa:2007zk,Miura:2018dsy} with a magnetic field added, 
which would be worth pursuing elsewhere. 

\item 
\textcolor{black}{It is of interest to clarify the full phase structure of the extended Columbia plot  
in three dimensions spanned by the $(m_0/f_\pi, m_s/f_\pi, eB/f_\pi^2)$ 
parameter space (such as an extension of Figs.~\ref{11} and~\ref{13}), 
whichever the approach goes beyond the mean 
field approximation, or not. 
}
\end{itemize} 

\acknowledgments
We thank Lang Yu for useful comments. 
This work was supported in part by the National Science Foundation of China (NSFC) under Grant No.11747308, 11975108, 12047569, 
and the Seeds Funding of Jilin University (S.M.).  
The work by M.K. was supported by the Fundamental Research Funds for the Central Universities.

\appendix 

\section{NJL formulas at vacuum}
\label{App:formulas}

In this Appendix we present a list on the QCD hadronic observables in the framework of the NJL model in Eq.(\ref{Lag:NJL})  
to get the QCD-monitor parameter setting in Eqs.(\ref{parameters1}) and (\ref{parameters2}). 
The details on the formulas to be given here can also be found in Refs.~\cite{Hatsuda:1994pi} and~\cite{Rehberg:1995kh}. 
The divergences arising in the loop functions are regularized 
by the three-momentum cutoff $\Lambda$ following the regularization scheme 
applied in the literature. 
For the four-momentum (Lorentz covariant) cutoff scheme, see, e.g., 
a review~\cite{Klevansky:1992qe}.

	\begin{itemize}
	\item The pion decay constant $f_{\pi}$: it is computed by directly evaluating quark loop contributions to the spontaneously broken $SU(2)$ axial current $J_\mu^{A, a}= \bar{l} \gamma_\mu \gamma_5 (\sigma^a/2) l$ ($a=1,2,3$),  
	through the definition of $f_\pi$, $\langle 0| J_\mu^{A \, a}(0) | \pi^b(p) \rangle  = - i p_\mu f_\pi \delta^{ab}$, where $l = (u,d)^T$.  
	In the NJL model with 
	the large $N_c$ limit taken (summing up the ring diagrams), we thus have 
		
		\begin{equation}
			f_{\pi} =G_{\pi q}(m_{\pi})M_{u}\frac{2 N_c}{\pi^2}\int^{\Lambda}_{0} dp \frac{p^2}{\sqrt{M_u^2 + p^2}[4(M_u^2+p^2)-m_{\pi}^2]}
			\,, 
			\end{equation}
		where $G_{\pi q}(p=m_{\pi})$ is the pion wavefunction renormalization amplitude evaluated at the onshell, 
		\begin{equation}
			G^2_{\pi q}(m_{\pi})=\left (\frac{N_c}{2\pi^2} \int^{\Lambda}_0 p^2 dp \frac{\sqrt{M_u^2+p^2}}{(p^2+M_u^2-\frac{m_{\pi}^2}{4})^2} \right )^{-1}
		\,. 
		\end{equation}

		\item The pion mass $m_\pi$ is computed by extracting the pole of the pion propagator dynamically generated by the quark loop contribution in the NJL with a resummation technique (Random Phase Approximation) applied~\cite{Hatsuda:1994pi}. The pole position is thus detected as 
		\begin{equation}
			1+G_{\pi}\Pi_{\pi}(m_\pi^2)=0
		\,,  
		\end{equation}
		where 
		\begin{align} 
		G_{\pi} &= 2 G -K \phi_s 
		\,, \notag\\ 
		\Pi_\pi &= -2\frac{N_c}{\pi ^2}\int^\Lambda _0  d p\, p^2  \frac{1}{\sqrt{p^2 + M_u^2}} 
		\,. 
		\end{align}
		This pion mass is actually related to the light quark condensates $\phi_u$ and $\phi_d$, via the 
		the low energy theorem (the so-called Gell-Mann-Oakes-Renner relation): 
		\begin{equation}
			m_{\pi}^2 =-m_0 (\phi_u + \phi_d)/f_{\pi}^2 = 
			- 2 m_0 \phi_u/f_{\pi}^2
			\,.  
			\end{equation}
		
	\item The kaon mass $m_{K}$ is calculable in the same way as in the case of $m_\pi$ above: 
	
		\begin{equation}
			1+G_{K}\Pi_{K}(m_K^2)=0
		\,, 
		\end{equation}
		where
		\begin{align}
			G_K &=2 G - K \phi_u
		\,, \notag\\ 
			\Pi_K(w) &=2F(w; u,s)+2F(-w; s, u)
		\,, \notag\\ 
			F(w; i,j) &=-\frac{N_c}{4\pi^2}\int^{\Lambda}_0 p^2 dp \left [ \frac{1}{E_{i}}f_{ij}(w) + \frac{1}{E_{j}}f_{ji}(w) \right ] 
		\,, \notag\\ 
		f_{ij}(w) &=2\frac{M_i(M_j-M_i)-E_iw}{E_j^2-(E_j+w)^2}
		\,,   
		\end{align}
with $E_i = \sqrt{p^2 + M_i^2}$. 

	\item The $\eta'$ mass $m_{\eta^{\prime}}$ is identified as the highest mass eigenvalue arising 
	from the mass mixing in the $0-8$ channel in terms of the basis of $U(3)$ generators. 
	Similarly to the pion and kaon cases, 
	$m_{\eta'}$ is then extracted as the highest pole for the mixed propagator in the $0-8$ channel, $D(q^2)$ as  
		
		\begin{equation}
		D_{\eta^{\prime}}(m_{\eta^{\prime}}^2)=0
		\,. 
		\end{equation}
		The mixed propagator $D(q^2)$ is given by 
		\begin{equation}
			D(q^2)=-G_P^{-1}\left ( \frac{1}{1+G_P \Pi_P(q^2)} \right ) \equiv \begin{pmatrix}
						A(q^2) & B(q^2)\\
						B(q^2) & A(q^2)
				\end{pmatrix}		
\,, 
		\end{equation}
with 
\begin{align} 
G_P &=
\begin{pmatrix}
	G_P^{00} & G_P^{08}\\
	G_P^{80} & G_P^{88}
\end{pmatrix}
=
\begin{pmatrix}
	2 G+ \frac{2}{3}(\phi_u +\phi_d +\phi_s)K &  
	\frac{\sqrt{2}}{6} (2\phi_s-\phi_u-\phi_d )K \\
	\frac{\sqrt{2}}{6} (2\phi_s-\phi_u-\phi_d )K & 
	2 G + \frac{1}{3}(\phi_s-2\phi_u -2\phi_d )K
\end{pmatrix}
\,, \notag\\ 
\Pi_P &=
\begin{pmatrix}
	\Pi_P^{00} & \Pi_P^{08}\\
	\Pi_P^{80} & \Pi_P^{88}
\end{pmatrix}
=
\begin{pmatrix}
	\frac{2}{3}	(2I_P^{uu}+I_P^{ss}) &  \frac{2\sqrt{2}}{3}(I_P^{uu}-I_P^{ss}) \\\frac{2\sqrt{2}}{3}(I_P^{uu}-I_P^{ss} ) & \frac{2}{3} (I_P^{uu} + 2I_P^{ss})
\end{pmatrix}
\,, \notag\\ 
I_P^{ii} & \equiv  - \frac{N_c}{\pi ^2}\int^\Lambda _0  d p\, p^2  \frac{1}{\sqrt{p^2 + M_i^2}}  
\,. 
\end{align}  
Through the diagonalization process like 
\begin{align} 
	T_{\theta}D^{-1}(q^2)T_{\theta}^{-1} &=
	\begin{pmatrix}
		D_{\eta^{\prime}}^{-1}(q^2) & 0 \\
		0 & D_{\eta}^{-1}(q^2)
	\,\end{pmatrix}
	\,, 
	\notag\\ 
{\rm with} \qquad	T_{\theta} &=
	\begin{pmatrix}
		\cos \theta & -\sin \theta \\
		\sin \theta & \cos \theta
	\end{pmatrix}
	\,, \qquad {\rm and} \quad \tan 2\theta =\frac{2B(q^2)}{C(q^2)- A(q^2)}
\,,  
\end{align}
we thus get $m_{\eta'}$ and also $m_\eta$.

\end{itemize}

\bibliographystyle{JHEP}
\bibliography{references}{}

\end{document}